\documentclass[preprint,aps,showpacs]{revtex4}

\usepackage{graphicx}
\usepackage{bm}       
\usepackage{mathptmx} 

\begin{document}

\title{Chiral anomalies and rooted staggered fermions}
\author{Michael Creutz}
\affiliation{
Physics Department, Brookhaven National Laboratory\\
Upton, NY 11973, USA
}

\begin{abstract}
A popular approximation in lattice gauge theory is an extrapolation in
the number of fermion species away from the four fold degeneracy
natural with the staggered fermion formulation.  I show that the
procedure mutilates the expected continuum holomorphic behavior in the
quark masses.  This is due to a chiral symmetry group that is of a
higher rank than desired.  The conventional resolution proposes
canceling the unphysical singularities with a plethora of extra states
appearing at finite lattice spacing.  This unproven conjecture
requires an explicit loss of unitarity and locality.  Even if correct,
the approach implies large cutoff effects in the low-energy
flavor-neutral sector.
\end{abstract}
\maketitle

Lattice gauge theory provides a powerful tool for the investigation of
non-perturbative phenomena in strongly coupled field theories, such as
the quark confining dynamics of the strong interactions.  However
numerical calculations are quite computer intensive, strongly
motivating approximations that reduce this need.  One such, the
valence or quenched approximation
\cite{Weingarten:1980hx,Hamber:1981zn}, introduces rather uncontrolled
uncertainties, but with the growth in computer power, its use is
currently being eliminated.

Another popular approximation \cite{Aubin:2004ej,Aubin:2004fs} arises
from the simplicity of the staggered fermion formulation \cite{
Kogut:1974ag,Susskind:1976jm,Sharatchandra:1981si}.  With only one
Dirac component on each site, the large matrix inversions required are
substantially faster than with other fermion formulations.  However
the approach and its generalizations are based on a discretization
method that inherently requires a multiple of four fundamental
fermions.  The reasons for this are related to the cancellation of
chiral anomalies.  To apply the technique to the physical situation of
two light and one intermediate mass quark requires an extrapolation
down in the number of fermions.  As usually implemented, the approach
involves taking a root of the fermion determinant inside standard
hybrid Monte Carlo simulation algorithms.  This step has not been
justified theoretically.  The purpose of this note is to show that at
finite lattice spacing this reduction inherently mutilates the quark
mass dependence expected in the continuum theory.  A preliminary
discussion of these points appears in Ref.~\cite{Creutz:2006ys}.

The method has its roots in the ``naive'' discretization of the
derivatives in the Dirac equation
\begin{equation}
\overline\psi \gamma_\mu \partial_\mu \psi \rightarrow
{1\over 2 a} \overline\psi_x \gamma_\mu (\psi_{x+ae_\mu}-\psi_{x-ae_\mu})
\end{equation}
with $a$ denoting the lattice spacing.  Fourier transforming to
momentum space, the momentum becomes a trigonometric function
\begin{equation}p_\mu\rightarrow {1\over 2 i a}
(e^{iap_\mu}-e^{-iap_\mu})={1\over a}\sin(ap_\mu)
\end{equation}
The natural range of momentum is $-\pi/a <p_\mu \le \pi/a$.  The
doubling issue is that the propagator has poles not just at small
momentum, but also when any component is near $\pi$ in magnitude.
These all contribute as intermediate states in Feynman diagrams; so,
the theory effectively has $2^4=16$ fermions.  I refer to these
multiple states as ``doublers'' or ``flavors'' in the following
discussion.

Note that the slope of the sine function at $\pi$ is opposite to that
at 0.  This can be absorbed by changing the sign of the corresponding
gamma matrix.  This changes the sign of $\gamma_5$ as well; so, the
doublers divide into different chirality subsets.  The determinant of
the Dirac operator is not simply the sixteenth power of a single
determinant.

Without a mass, the naive action has an exact chiral symmetry of the
kinetic term under
\begin{eqnarray}
&\psi\rightarrow e^{i\theta\gamma_5}\psi\cr
&\overline\psi\rightarrow \overline\psi e^{i\theta\gamma_5}
\label{chiral}
\end{eqnarray}
The conventional mass term is not invariant under this rotation
\begin{equation}
m\overline\psi \psi\rightarrow m\overline\psi \psi \cos(2\theta)
+im\overline\psi\gamma_5 \psi\sin(2\theta)
\end{equation} 
Thus any mass term of the form on the right hand side of this relation
can have theta rotated away.  This is consistent with known anomalies
since this is in reality a flavor non-singlet chiral rotation.  The
different species use different signs for $\gamma_5$.  As special
cases, in this theory $m,-m,$ and $\pm i\gamma_5 m$ are all physically
equivalent.

To arrive at the staggered formulation, note that whenever a fermion
hops between neighboring sites in direction $\mu$, it picks up a
factor of $\gamma_\mu$.  An arbitrary closed fermion loop on a
hyper-cubic lattice gives a product of many gamma factors, but any
particular component always appears an even number of times.  Bringing
them through each other using anti-commutation, the net factor for any
loop is proportional to unity.  Gauge fields don't change this fact
since they just involve $SU(3)$ phases on the links.  So if a fermion
starts in one spinor component, it returns to the same component after
the loop.  The 4 Dirac components give 4 independent theories.  There
is an exact $SU(4)$ symmetry.  Without a mass term, this is actually
an exact $SU(4)\otimes SU(4)$ chiral symmetry \cite{Karsten:1980wd}.

Staggered fermions single out one component on each site.  Which
component depends on the gamma factors to get to the site in question
from one chosen starting site.  Ignoring the other components reduces
the degeneracy from 16 to 4.  The process brings in various
oscillating phases from the gamma matrix components.  One explicit
projection that accomplishes this is (using integer coordinates and
the convention $\gamma_5=-\gamma_1\gamma_2\gamma_3\gamma_4$ with
Euclidean gamma matrices)
\begin{equation}
P=P^2={1\over 4} \left(1+i\gamma_1\gamma_2 (-1)^{x_1+x_2}
+i\gamma_3\gamma_4
(-1)^{x_3+x_4}+\gamma_5(-1)^{x_1+x_2+x_3+x_4}\right)
\label{projection}
\end{equation} 
Note that some degeneracy must remain.  No chiral breaking appears in
the action, and all infinities are removed.  The conventional axial
anomaly is canceled between the remaining species.  Furthermore, the
naive replacement $\psi\rightarrow\gamma_5\psi$ exactly relates the
theory with mass $m$ and mass $-m$.  With 4 flavors this symmetry is
allowed since it still represents a flavored chiral rotation.  The
doublers appear in chiral pairs.

To proceed I sketch how a typical simulation with fermions proceeds.
For a generic fermion matrix $D$, the goal of a the simulation is to
generate configurations of gauge fields $A$ with a probability
\begin{equation}
P(A)\propto \exp(-S_g(A)+N_f {\rm Tr}\ \log(D(A)))
\end{equation}
Here $S_g$ is the pure gauge part of the action and $N_f$ is the
number of fermion species.  With some algorithms additional commuting
``pseudo-fermion'' fields are introduced
\cite{Fucito:1980fh,Scalapino:1981qs}, but these details are not
important to the following discussion.  With staggered or naive
fermions the eigenvalues of $D$ all appear in complex conjugate pairs;
thus, the determinant is non-negative as necessary for a probability
density.

In hybrid Monte Carlo schemes \cite{Duane:1987de} auxiliary
``momentum'' variables $P$ are introduced, one for each degree of
freedom in $A$.  The above distribution is generalized into
\begin{equation}
P(A,P)\propto \exp\left(-S_g(A)+N_f {\rm Tr}\ \log(D(A))+\sum
P_i^2/2\right)
\label{pap}
\end{equation} 
As the momenta are Gaussian random variables, it is easy to generate a
new set at any time.  For the gauge fields one sets up a
``trajectory'' in a fictitious ``Monte Carlo'' time variable $\tau$
and uses the exponent in (\ref{pap}) as a classical Hamiltonian
\begin{equation}
H=\sum P_i^2/2+V(A)
\end{equation}
with the ``potential''
\begin{equation}
V(A)=-S_g(A)+N_f {\rm Tr}\log(D(A)).
\end{equation}
The Hamiltonian dynamics
\begin{eqnarray}
&{dA_i\over d\tau}=P_i\cr
&{dP_i\over d\tau}= F_i(A)=-{\partial V(A)\over \partial A}
\end{eqnarray}
conserves energy and phase space.  Under such evolution the
equilibrium ensemble stays in equilibrium, a sufficient condition for
a valid Monte Carlo algorithm.  After evolution along a trajectory of
some length $\tau$, discretized time steps $\delta\tau$ can introduce
finite step errors and give a small change in the ``energy.''  The
hybrid Monte Carlo algorithm corrects for this with a Metropolis
accept/reject step on the entire the trajectory.  The trajectory
length and step size are parameters to be adjusted for reasonable
acceptance.  After the trajectory one can refresh the momenta by
generating a new set of gaussianly distributed random numbers.  The
procedure requires the ``force'' term
\begin{equation}
 F_i(A)=-{\partial V(A)\over \partial A}
={\partial S_g(A)\over \partial A}-N_f{\rm Tr} 
\left( D^{-1} {\partial D(A)\over \partial A}\right).
\label{force}
\end{equation}
To calculate the second term requires an inversion of the sparse
matrix $D$ applied to a fixed vector.  Standard linear algebra
techniques such as a conjugate gradient algorithm can accomplish this.
In practice this step is the most time consuming part of the
algorithm.

Returning to staggered fermions, one would like to eliminate the
unwanted degeneracy by a factor of four.  One attempt to do this
reduction involves an extrapolation in the number of flavors.  In the
molecular dynamics trajectories for the simulation of the gauge field,
the coefficient of the fermionic force term in Eq.~(\ref{force}) is
arbitrarily reduced from $N_f$ to $N_f/4$, where $N_f$ is the desired
number of physical flavors.  Although not proven, this seems
reasonable when $N_f$ is itself a multiple of four.  The controversy
arises for other values of $N_f$.

Here I argue that the procedure is an approximation that inherently
mutilates the analytic structure expected in the quark masses.  To see
this consider the case of two flavor QCD with quark masses $m_u$ and
$m_d$.  Complexifying the mass terms in the usual way
\begin{equation}
\sum_{a=u,d}
{\rm Re}\ m_a\ \overline\psi^a\psi^a
 +i\ {\rm Im}\ m_a\ \overline\psi^a\gamma_5\psi^a
\end{equation}
the physical theory is invariant under the flavored chiral rotation
\begin{eqnarray}
&m_u\rightarrow e^{i\theta}m_u\cr
&m_d\rightarrow e^{-i\theta}m_d
\end{eqnarray}
Due to the chiral anomaly, it must not be invariant under the singlet
chiral rotation
\begin{eqnarray}
&m_u\rightarrow e^{i\theta}m_u\cr
&m_d\rightarrow e^{i\theta}m_d
\end{eqnarray}
The symmetry in mass parameter space requires that the rotations of
the up and down quark masses be in opposite directions.  Indeed, this
limitation is correlated with there only being one neutral Goldstone
boson for the two flavor theory.

Now formulate this theory with two independent staggered fermions, one
for the up and one for the down quark, each reduced using the rooting
procedure.  From Eq.~(\ref{projection}), the corresponding
complexification of the staggered mass term takes the form
\begin{equation}
\sum_{a=u,d} ({\rm Re}\ m_a + i S(j)\ {\rm Im}\ m_a)\ \psi^\dagger(j)\psi(j)
\end{equation}
with $S(j)$ being $\pm 1$ depending on the parity of the site $j$.
The issue arises from the fact that that the staggered fermion
determinant, and therefore the path integral, are exactly invariant
under $m\rightarrow e^{i\theta}m$ for either the up or the down quark.
This is too much symmetry in parameter space.  The physical $SU(2)$
chiral symmetry group is of rank one, while the chiral symmetry of the
two flavored staggered fermion formulation has rank two.  It requires
two neutral Goldstone bosons in the massless limit, rather than the
one of the physical theory.

The issues become particularly severe in the chiral limit when when
$N_f$ is odd.  For the staggered theory, the fermion determinant is a
function of $m^2$.  The surviving chiral symmetry gives equivalent
physics for either $m$ or $-m$.  However, it is well known that with
an odd number of flavors, physics has no symmetry under changing the
sign of the mass \cite{DiVecchia:1980ve, Creutz:1995wf,
Creutz:2003xu}.  The most dramatic demonstration of this appears in
the one flavor theory \cite{Creutz:2006ts}.  In this case anomalies
break all chiral symmetries and no Goldstone bosons are expected.  The
theory behaves smoothly as the mass parameter passes through zero.
The lightest meson, call it the $\eta^\prime$, acquires a mass through
anomaly effects, and the lowest order quark mass corrections are
linear
\begin{equation}
m_{\eta^\prime}^2(m)=m_{\eta^\prime}^2(0)+ c m
\label{etaprime}
\end{equation} 
Such a linear dependence in a physical observable is immediately
inconsistent with $m\leftrightarrow -m$ symmetry.

The one flavor case is perhaps a bit special, but there are similar
problems with the three flavor situation \cite{Creutz:2003xu}.
Identify the quark bi-linear with an effective chiral field
$\overline\psi_a\psi_b\sim \Sigma_{ab}.$ Here $a$ and $b$ are flavor
indices.  The $SU(3)\otimes SU(3)$ chiral symmetry of the massless
theory is embodied in the transformation
\begin{equation}
\Sigma \rightarrow g_L^\dagger \Sigma g_R
\end{equation}
with $g_L,g_R \in SU(3)$.  For positive mass, $\Sigma$ should have an
expectation value proportional to the $SU(3)$ identity $I$.  This is
not equivalent to the negative mass theory because $-I$ is not in
SU(3).  Indeed, for negative mass it is expected that the infinite
volume theory spontaneously breaks $CP$ symmetry, with
$\langle\Sigma\rangle\propto e^{\pm 2\pi i/3}$
\cite{Montvay:1999gn,Creutz:2003xu}.

These qualitative effective Lagrangian arguments are quite powerful
and general.  Another way to see the one flavor behavior is to start
with a larger number of flavors, say 3 or 4, and make the masses
non-degenerate.  As only one of the masses passes through zero, the
behavior for the lightest meson mimics that in Eq.~(\ref{etaprime}).
Extrapolated staggered quarks with their symmetry under taking any
quark mass to its negative will miss the linear term.

Small real eigenvalues of the Dirac operator are responsible for these
effects.  The odd terms come from topological structures in the gauge
fields \cite{'tHooft:fv}.  For small mass in the traditional continuum
discussion, $|D|\sim m^\nu$ with $\nu$ the winding number of the gauge
field.  The condensate
\begin{equation}
\langle\overline\psi \psi\rangle=
{1\over Z}\int (dA) |D|^{N_f} e^{-S_g(A)}\ {\rm Tr}\ D^{-1}
\end{equation}
receives a contribution going as $m^{N_f-1}$ from the $\nu=1$ sector.
For the one flavor case, this is an additive constant.  This constant
will be missing from the extrapolated staggered theory because of the
symmetry in Eq.~(\ref{chiral}).  This phenomenon is also responsible
for the fact that a single massless quark is not a well defined
concept \cite{Creutz:2004fi}.

For the general odd flavor case, the odd winding number terms have the
opposite symmetry under the sign of the mass than the even terms,
although with more flavors this starts at a higher order in the mass.
For 3 flavors the condensate at finite volume will display a $m^2$
correction to the leading linear behavior.  The extrapolation down
from the staggered 4 flavor theory will not see this term.

It is during the transitions between topological sectors that the
unrooted theory behaves quite differently than the target theory.  In
particular, with a smooth gauge field of unit winding number near the
continuum limit, the unrooted theory should have four small
eigenvalues representing the zero modes from the index theorem.
Considering an evolution of the gauge fields from zero to unit winding
number, two of these drop down from positive imaginary part and two
move up from below.  Any approximate four fold degeneracies between
the higher eigenvalues must break down during this evolution.
Attempts to define the rooting procedure by selecting one fourth of
the eigenvectors will necessarily involve ambiguities.

While I have shown diseases with the chiral behavior of extrapolated
staggered fermions at finite cutoff, it has been suggested
\cite{Durr:2004ta,Bernard:2006vv,Durr:2006ze} that these problems go
away as the cutoff is removed.  Indeed, in quantum field theory we are
accustomed to the non-commutation of certain limits, such as vanishing
mass and infinite volume when a symmetry is being spontaneously
broken.  In that case the mass and the volume are both infrared
issues.  As the lattice is an ultraviolet regulator and the chiral
issues raised here involve long distance physics, it seems peculiar
for the order of these limits to affect each other.  Nevertheless,
suppose that taking the cutoff to zero before taking the massless
limit does give the correct physics.  Then the regulator must
introduce singularities that are not present in the continuum theory.

The issue is again clearest for the one flavor theory, where in the
continuum the condensate, $\langle \overline\psi\psi \rangle$
appropriately renormalized, does not vanish and is smoothly behaved
around m=0.  Analyticity in the mass is expected with a radius of
order the eta-prime mass-squared over the typical scale of the strong
interactions, $\Lambda_{\rm qcd}$.  Now turn on the extrapolated
staggered regulator.  At $m=0$, $\langle \overline\psi\psi \rangle$
must suddenly jump to zero.  For every eigenvalue of the staggered
fermion matrix at vanishing mass, its negative is also an eigenvalue.
Thus configuration by configuration the trace of $D^{-1}$, and thus
the condensate, is incorrectly predicted to be identically zero.
Furthermore, due to confinement and the chiral anomaly, this
unphysical jump occurs both at finite volume and in the absence of any
massless physical particles for the continuum theory.

This problem generalizes to the multi-flavor theory with non-degenerate
quark masses.  The proposed regulator forces the condensate associated
with any given species to vanish with the corresponding mass, in direct
contradiction with the continuum behavior expected from effective
Lagrangian analysis.  Physical observables at specific points in
parameter space where continuum physics is smooth are forced to
develop infinite derivatives with respect to the cutoff as it is
removed.  Even if this occurs only in the vicinity of isolated points,
this seems an absurd behavior for an ultraviolet regulator and is in
strong contrast to more sensible schemes such as Wilson fermions
\cite{Wilson:1975id}.

It has recently been argued \cite{Bernard:2006vv} that this unphysical
behavior could be avoided in the continuum limit as long as one stays
away from these singularities.  Consider the two flavor theory
discussed earlier.  Due to the doubling, the unrooted theory has 32
neutral pseudo-scalar mesons.  The anomaly should give one of these a
mass of order the QCD scale, and this becomes the eta prime.  At
finite lattice spacing the remaining 31 particles divide into two
exact Goldstone bosons corresponding to the exact chiral symmetries
and 29 approximate Goldstone bosons.  If we now give only one of the
quarks a small mass, one of the massless pseudo-scalars should acquire
a mass and represent the neutral pion.  The second, however, must
remain massless due to the remaining symmetry.
Ref.~\cite{Bernard:2006vv} argues that at finite lattice spacing the
29 extra mesons at finite mass are still there after rooting.  They
suggest, without proof, that if the second quark is given a small mass
and as the lattice spacing is taken zero, it is possible that this
plethora of extra states could move down in energy and cancel the
unwanted extra Goldstone boson.  This scheme requires a loss of
unitarity; indeed, the production cross sections for some pairs of the
unwanted mesons must be negative so the total production can add to
zero.  And before this happens the theory is non-local because of long
range forces due to the one unwanted massless particle.

Such a mechanism appears to me as rather contrived, but
Ref.~\cite{Bernard:2006vv} suggests that it is merely an ugly feature
of the algorithm.  Even if the proposed cancellation does occur, at
finite lattice spacing we have a factor of 16 more neutral
pseudo-scalar mesons than in the physical theory.  This suggests that
the lattice corrections to physics in the flavor singlet sector ar
potentially rather large.

To summarize, at finite lattice spacing the holomorphic behavior in
the quark masses for rooted staggered quarks is qualitatively
incompatible with continuum physics.  The chiral symmetry group with
rooted fermions is of a higher rank than desired.  This gives rise to
unphysical singularities when any single quark mass vanishes.  For the
extra symmetry to disappear as the lattice spacing is taken to zero
requires rather subtle cancellations which have not been demonstrated.
The approximation may still be reasonable for some observables, most
particularly those involving only flavor non-singlet particles.  But
any predictions for which anomalies are important are particularly
suspect.  This includes the $\eta^\prime$ mass, but also more mundane
quantities such as the lightest baryon mass, which in the chiral limit
is entirely non-perturbative.

\section*{Acknowledgments}
This manuscript has been authored under contract number
DE-AC02-98CH10886 with the U.S.~Department of Energy.  Accordingly,
the U.S. Government retains a non-exclusive, royalty-free license to
publish or reproduce the published form of this contribution, or allow
others to do so, for U.S.~Government purposes.

\end{document}